\documentclass[aps,prl,reprint,10pt,longbibliography,superscriptaddress,nofootinbib]{revtex4-2}

\usepackage{amsmath, amssymb, amsfonts, amsthm, bm, mathrsfs, bbm, enumerate,mathtools}
\usepackage{graphicx}
\usepackage{epstopdf}
\usepackage{color}
\usepackage[usenames,dvipsnames]{xcolor}
\usepackage[citecolor=blue, colorlinks=true, urlcolor=blue, linkcolor=blue]{hyperref}
\usepackage{bbold}
\usepackage{feynmp-auto} %for feynman diagrams

\usepackage{soul}
\usepackage{float}

\newcommand{\eq}[1]{Eq.~(\ref{#1})}

\usepackage{color}
\usepackage{nicefrac}
\usepackage{dsfont} %for ensemble alphabet
\usepackage{wasysym} %for hexagon (in Eq environment or in text)
\usepackage{stmaryrd} % for [| and |] brackets, use \llbracket and \rrbracket
\usepackage{hyperref} %for links in pdf
\usepackage{xcolor}
\usepackage{empheq} %for e.g. boxing align environment

\definecolor{airforceblue}{rgb}{0.36, 0.54, 0.66}

\usepackage{verbatim}

\usepackage{cancel}
\usepackage{url}

\newcommand{\mb}{\mathbf}
\newcommand{\bs}{\boldsymbol}

\newcommand{\mc}{\mathcal}

\newcommand\varpm{\mathbin{\vcenter{\hbox{
  \oalign{\hfil$\scriptstyle+$\hfil\cr
          \noalign{\kern-.3ex}
          $\scriptscriptstyle({-})$\cr}
}}}}

\begin{document}

\newcommand{\ourtitle}{Tuning transport in solid-state Bose-Fermi mixtures by Feshbach resonances}
\title{\ourtitle}

\newcommand{\TUM}{\affiliation{Technical University of Munich, TUM School of Natural Sciences, Physics Department, 85748 Garching, Germany}}
\newcommand{\MCQST}{\affiliation{Munich Center for Quantum Science and Technology (MCQST), Schellingstr. 4, 80799 M{\"u}nchen, Germany}}
\newcommand{\Harvard}
{\affiliation{Department of Physics, Harvard University, Cambridge, Massachusetts 02138, USA}}

\author{Caterina Zerba} 
\thanks{These authors contributed equally to this work.\\}
\TUM 
\MCQST 

\author{Clemens Kuhlenkamp} 
\thanks{These authors contributed equally to this work.\\}
\Harvard 
\TUM 
\MCQST 

\author{L\'eo Mangeolle} 
\thanks{These authors contributed equally to this work.\\} 
\TUM 
\MCQST

\author{Michael Knap} 
\TUM 
\MCQST

\begin{abstract}

Transition metal dichalcogenide (TMD) heterostructures have emerged as promising platforms for realizing tunable Bose-Fermi mixtures. Their constituents are fermionic charge carriers resonantly coupled  to long-lived bosonic interlayer excitons, allowing them to form trion bound states.
Such platforms promise to achieve comparable densities of fermions and bosons at low relative temperatures.
Here, we predict the transport properties of correlated Bose-Fermi mixtures close to a narrow solid-state Feshbach resonance. 
When driving a hole current, the response of doped holes, excitons, and trions are significantly modified by the resonant interactions, leading to 
deviations from the typical Drude behavior and to a sign change of the exciton drag. Our results on the temperature-dependent resistivities demonstrate that near resonance interaction effects dominate over established conventional scattering mechanisms in these solid-state Bose-Fermi mixtures.

\end{abstract}

\date{\today}

\maketitle

Unconventional phases in solids are predicted to arise when a Fermi surface is strongly coupled to bosonic excitations, such as phonons, spin- and density-wave fluctuations, and collective modes emerging in the vicinity of phase transitions~\cite{Landau1933,Pekar1946,Hertz1976,Lee2006,Lhneysen2007,Sachdev2011,Li_2021,Senthil2008}.  
However, isolating relevant interaction channels is challenging, as electrons are typically coupled simultaneously to multiple bosonic modes. This motivates the exploration of Bose-Fermi mixtures in more controlled settings of transition-metal-dichalcogenide (TMD) heterostructures~\cite{Ma2021,schwartz21,Park2023,Xiong2023,Gao2024,Lian2024,Nguyen23,Qi2023,upadhyay2024giantenhancementexcitondiffusion, mhenni2024gatetunablebosefermimixturestrongly}; 
complementary regimes are accessible in ultracold atomic gases as well~\cite{Chin2010,zwerger_rev_08,barbut15,DeSalvo2019,Yan24,Duda2023}.
TMDs offer the advantage that low relative temperatures are reachable~\cite{Nguyen23}. In these settings, fermions are introduced by charge doping and high densities of long-lived bosons are realized as tightly-bound interlayer excitons~\cite{Wilson2021}. Excitons interact with doped charges~\cite{sidler_polaron,schwartz21,Fey2020,Kuhlenkamp21} forming fermionic bound states, referred to as trions, which have been observed to remain stable even at finite densities~\cite{Nguyen23,Qi2023}. While theoretical studies have shown that low temperature Bose-Fermi mixtures in TMDs could host sought-after unconventional phases~\cite{Zerba2024,vonMilczewski2024, kumar2024, crepel2023}, surprisingly many properties of such mixtures at intermediate temperatures remain unexplored. 
Recent experimental advances ~\cite{Joe2024, Pack2024Jul} are furthermore enabling new theoretical and experimental avenues for studying correlated Bose-Fermi mixtures through transport; an approach that is very challenging in cold atom setups. 

In this work, we theoretically investigate transport properties of tunable solid-state Bose-Fermi mixtures in TMD heterostructures near a Feshbach resonance. In this regime, we map the strongly-interacting hole-exciton problem into an 
effective, analytically tractable theory that includes all three particle species: holes, excitons and trions. Our analysis shows how strongly interacting Bose-Fermi mixtures can be investigated in the vicinity of a solid-state Feshbach resonance. 
We uncover a rich phenomenology by considering a selectively driven hole current and tuning the exciton-hole scattering into resonance. Our results show that the conductivities of all three particle species depend sensitively on one external parameter, the perpendicular electric field, which tunes the relative energy of the trion. 
%In analogy with cold atomic gases, we exploit the control of exciton-charge scattering by a solid-state Feshbach resonance, and thus scattering depends sensitively on the trion energy which is tunable by a perpendicular electric field~\cite{Kuhlenkamp21,schwartz21,wagner2023}. 
Remarkably, this exciton-hole scattering effect dominates the transport properties below the phonon temperature scale, and induces a sign-changing exciton drag conductivity in a broad parameter regime. As another striking effect we find that the resistivity of the system exhibits a strong, non-monotonic temperature dependence, as well as an unconventional ac response beyond Drude phenomenology.

\begin{figure*}[!t]
\centering
\includegraphics[width=0.98\textwidth]{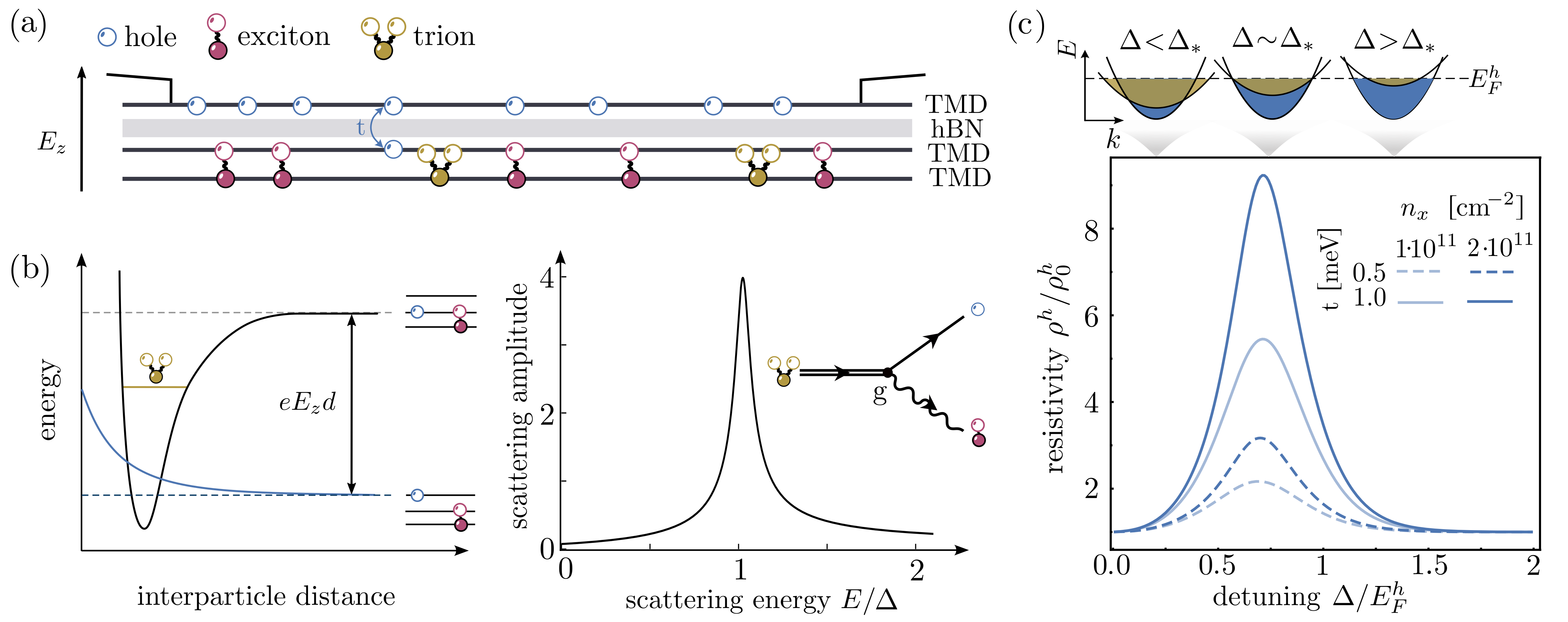}
\caption{\textbf{Setup, particle-like trion bound state and enhanced transport close to resonance.}
\textbf{a)} Proposed trilayer TMD structure. The system realizes a Bose-Fermi mixture with doped charges (holes) in the upper layer and strongly bound interlayer excitons in the lower two layers. A longitudinal electric field is applied only to the top layer via direct contacts that drive a hole current. 
\textbf{b)} Solid-state Feshbach resonance. 
Left: The energy of a spatially separated
exciton-hole pair (blue curve) 
can be tuned close to the energy of the trion (yellow) by the perpendicular electric field, which shifts the energy of the trion to $E_t^0 +\Delta = e E_z d$.
Right: The exciton-hole scattering amplitude $|f(E)|$ (see supplemental material~\cite{supp}) is resonantly enhanced around the shifted trion energy due to the narrow Feshbach resonance. 
\textbf{c)} Resistivity of holes in the upper layer $\rho^h$ as a function of the detuning $\Delta$, calculated at $T = 5.8$ K. 
The resistivity is strongly enhanced close to the resonance condition $\Delta \sim \Delta_\star$, where the hole and trion Fermi surfaces have the same size. Top: Sketches to illustrate the hole and trion Fermi seas as a function of $\Delta$.}
\label{fig:1}
\end{figure*}

\textbf{Model.---} Inspired by recent experiments~\cite{schwartz21,Nguyen23,Qi2023}, we propose a setting composed of three monolayer TMDs, where the top layer is hole-doped and separated from the other layers by hexagonal boron nitride (hBN) of thickness $d$, see Fig.~\ref{fig:1}~(a). The relative energy of charges in the middle layer is tuned by a perpendicular electric field $E_z$. This imposes an electrostatic potential difference $e E_z d$, where $e$ is the elementary charge. We consider fields $E_z$ that %increase the trion energy and 
suppress tunneling of unbound holes to the middle layer, see Fig.~\ref{fig:1}~(a,b). The perpendicular electric field and the density of doped charges are fixed independently by tuning the voltage of the top and middle layers, as shown in Ref.~\cite{schwartz21}. Interlayer excitons between the middle and the lower layer are optically or electrically injected, and the applied voltage does not relevantly affect their density since they are tightly bound \cite{schwartz21,Wang19}.
Interlayer excitons are selectively excited, as their binding energy is in general different from that of other excitations, e.g.\ intralayer excitons. %The density of excitons can also be tuned independently from the other parameters, when optically excited.
Due to the small overlap of the hole and electron wavefunctions, their lifetime can exceed hundreds of ns, which allows us to treat excitons as well-defined bosonic particles~\cite{Wilson2021}.

The Bose-Fermi mixture of excitons and holes is strongly interacting, which is reflected by the existence of a trion bound state. Scattering is tuned in resonance by changing the field $E_z$. It is possible to approximate multiple (resummed) scattering events between holes and excitons as an interaction mediated by the trion. This approximation, which is accurate near the Feshbach resonance, is discussed further in the Supplemental Material~\cite{supp}, and makes the transport problem analytically tractable. The bare hole-exciton-trion vertex $g$ gives the effective interaction between these three species, and is represented in Fig. ~\ref{fig:1}~(b). Starting from the microscopic exciton-hole Hamiltonian, we identify 
$g \simeq \left ( \frac{2\pi}{m_{\text{red}} |E_t^0|}\right )^{1/2} {\rm t}$, where $\rm t$ is the hole coherent tunneling rate, $E_t^0$ the trion binding energy, $m_{\text{red} }^{-1}= m_h^{-1} + m_x^{-1} = (3/2)\,m_h^{-1}$, and we have assumed that $g$ is momentum-independent~\cite{zwerger_rev_08,Zerba2024,Kuhlenkamp21}. For $e E_z d \geq E_t^0$, the trion decays into an exciton-hole pair at a rate proportional to $\rm t^2$, which can be much 
smaller than the relevant Fermi energies~\cite{schwartz21}. This allows us to treat trions as sharp quasi-particles whose energy is tuned by the Feshbach resonance, see Fig.~\ref{fig:1}~(b). Even for finite hole and exciton densities, the trion retains a quasi-particle peak, provided $g \ll E_F^h/\sqrt{n_x}$, which we use as a perturbative parameter.

The resulting system is well described by holes, excitons and trions with quadratic dispersions 
$\epsilon_{h,\mb p}= \frac{\mb p^2}{2m_h}-\mu_h$, $\omega_{\mb k}= \frac{\mb k^2}{2m_x}-\mu_x$
and $\epsilon_{t,\mb p}= \frac{\mb p^2}{2m_t}-\mu_t + \Delta$, respectively, 
where $\Delta= e E_z d -E_t^0$. The trion mass is $m_t = m_x + m_h$. We assume the system has reached thermal equilibrium, which fixes the chemical potentials to satisfy $\mu_h+\mu_x=\mu_t$~\cite{Powell05}.
The effective Hamiltonian describing the solid-state Bose-Fermi mixture is then
\begin{equation}
\begin{aligned}
 \hat{H}\!&=\!\! \int \!\!\!\tfrac{\mathrm{d}^2 \mb p}{(2\pi)^2} \left ( \epsilon_{h,\mb p} \,\hat{c}^\dagger_{\bold{p}} \hat{c}_{\bold{p}} 
 +  \epsilon_{t,\mb p}\, \hat{m}^\dagger_{\bold{p}} \hat{m}_{\bold{p}} \right )
 \!+\!\! \int \!\!\! \tfrac{\mathrm{d}^2 \mb k}{(2\pi)^2} \,\omega_{\mb k}\, \hat{x}^\dagger_{\bold{k}} \hat{x}_{\bold{k}} \\
 & +g  \int  \!\!\!\tfrac{\mathrm{d}^2 \mb p}{(2\pi)^2}\tfrac{\mathrm{d}^2\mb k}{(2\pi)^2} \left ( \hat{m}^\dagger_{\bold{p}+\bold{k}}\hat{c}_{\bold{p}}\hat{x}_{\bold{k}} + \text{h.c.} \right ),
 \label{eq:eff_model}
 \end{aligned}
\end{equation}
where $\hat{c}^\dagger, \hat{x}^\dagger$ and $\hat{m}^\dagger$ ($\hat{c}, \hat{x}$ and $\hat{m}$) are the hole, exciton and trion creation (annhilation) operators.  

Unless stated otherwise, we will use the following parameters: hole Fermi energy $E_{F}^h = 5$ meV, $E_t^0=-10$ meV, $\rm t=1$ meV, $n_x = 2 \cdot 10^{11}$ cm$^{-2}$ and $m_h = 0.5\,m_{\sf e}$ with $m_{\sf e}$ the free electron mass, and set $\hbar=1$~\cite{Kormanyos15,schwartz21,Jauregui2019Nov,Nguyen23,Qi2023}. Throughout the work we model 
extrinsic sources of momentum relaxation (e.g. disorder) for all species by a momentum-independent relaxation time $\tau_{0}= 10$~ps, motivated by recent transport experiments~\cite{Joe2024,Pack2024Jul,Guo2024}. The extrinsic sources of relaxation may include other excitations and few-body complexes, which are off-resonant and whose effect thus varies slowly with $E_z$ in the regime of interest. %As already realized experimentally,  perpendicular electric field, hole's density in the top layer and excitons' density, can be tuned independently, see Ref.~\cite{schwartz21}, and Ref. \cite{Wang19} for electrically injected excitons. 
We emphasize that our results do not depend qualitatively on the chosen parameters. %Furthermore, at low doping Wigner crystals the low doped charge density regime, where ordered phases such as Wigner crystals may form, can be easily avoided experimentally.

\textbf{Methods.---} Electrical contacts are placed such that a longitudinal electric field is applied only to the charge-doped layer, as shown in Fig.~\ref{fig:1}(a). This induces a charge current $j^h_a=\sigma^h_{ab} E_{b}$ in the upper layer, where $\sigma^h_{ab}$ is the hole conductivity tensor and $a,b \in \lbrace x,y\rbrace$ are the longitudinal and transverse directions. Since we are considering linear response in 
an isotropic and time-reversal symmetric model, 
$\sigma^h_{ab}= \delta_{ab}/\rho^h$ with $\rho^h$ the hole resistivity. 
Excitons and trions do not couple to the electric field directly. Instead, their particle currents, $J_a^x$ and $J_a^t$, induced in the middle and lower layers, arise purely from drag effects mediated by the hole current via many-body scattering. We define the corresponding conductivity as 
$\sigma^{i}_{ab}=eJ^{i}_a/E_b$ for $i\in\{x,t\}$.
We analyze the currents induced in the system for all three species of particles, as a function of detuning $\Delta$ and temperature $T$. Using a kinetic theory, where the collision integrals are obtained from a perturbative calculation of the self-energies~\cite{Kamenev2011Sep}, we derive and solve a set of three coupled Boltzmann's equations for the particle distributions. A perturbative calculation of the hole conductivity based on Kubo's formula~\cite{Mahan} yields similar results, up to quantitative corrections in the vicinity of the resonance. Further details about both methods, and a comparison, are reported in the Supplemental Material~\cite{supp}.

\textbf{Tunable hole transport.---}
\begin{figure}[t]
\centering
\includegraphics[width=0.437\textwidth]{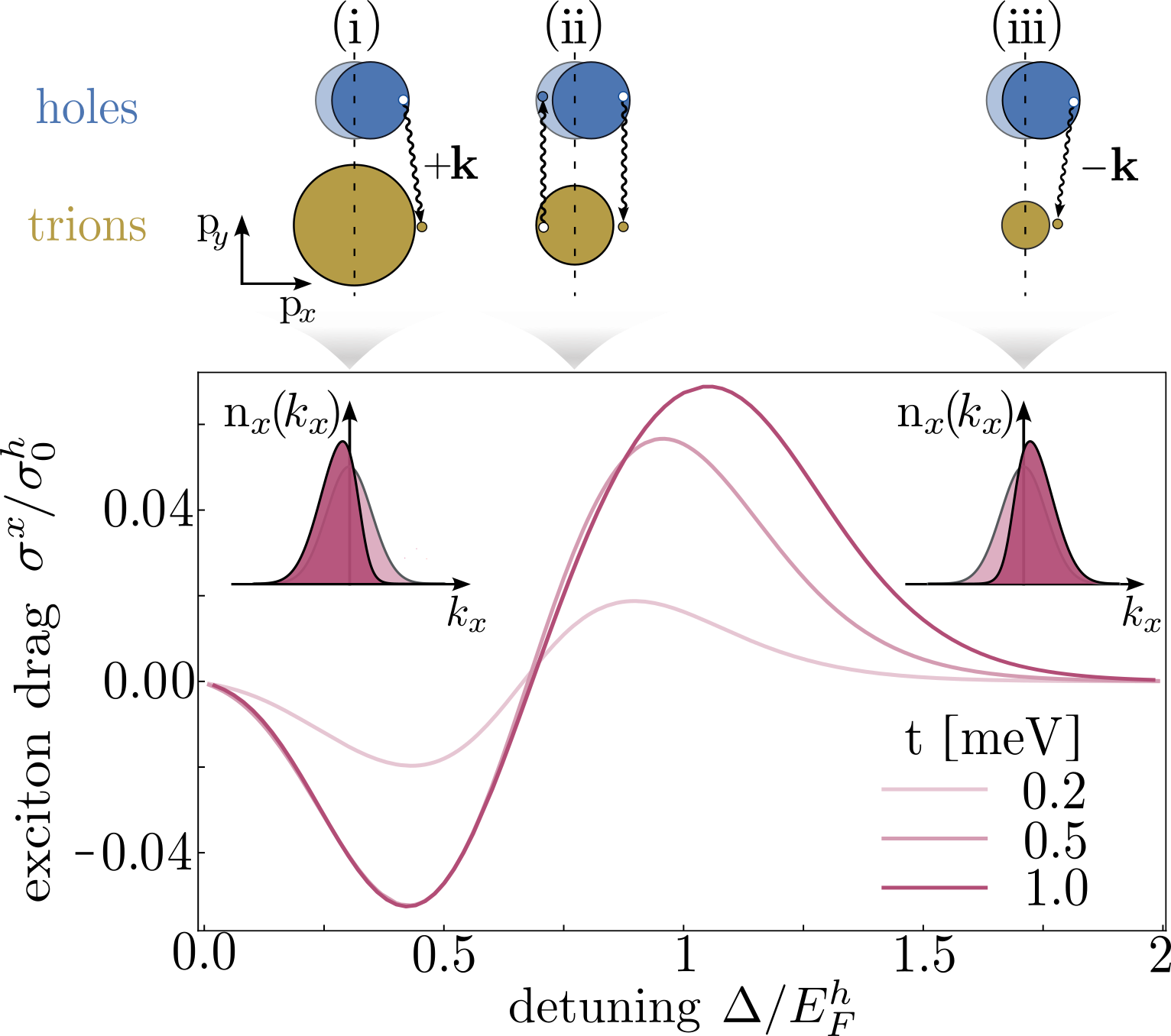}
\caption{\textbf{Sign-changing exciton drag.} Exciton drag conductivity as a function of detuning $\Delta$ evaluated  at temperature $T=5.8$ K.
The backflow of excitons, $\sigma^x<0$, for small detuning can be understood from the interplay of energy and momentum conservation and Pauli exclusion. Upper panels: Sketches of the relevant scattering processes for exciton drag.
(i) A hole (from the blue Fermi sea) absorbs a right-moving exciton (squiggly arrow) to form a trion (above the yellow Fermi sea). This depletes right-moving excitons and skews the exciton distribution $n_x(k_x)$ (red) against the direction of the hole current. 
(ii) Close to resonance, scattering processes primarily involve small-momentum excitons, leading to a small exciton drag.
(iii) A hole absorbs a left-moving exciton, which skews the exciton distribution  $n_x(k_x)$ in the direction of the hole current. }
\label{fig:2}
\end{figure}
In our system, holes exhibit a tunable resistivity $\rho^h$: the resistivity $\rho^h$ is governed not only by the intrinsic background scattering rate $\tau_{0}^{-1}$ that sets the constant background resistivity $\rho^h_0=2\pi/(\mu_h e^2\tau_{0})=1/\sigma_0^h$, but crucially also by many-body interactions with the bosonic interlayer excitons that depend on the perpendicular electric field $E_z=(\Delta+E_t^0)/ed$.
By solving the coupled transport equations, we find that the resistivity $\rho^h$ exhibits a strong resonant behaviour as a function of the detuning $\Delta$,
as demonstrated in Fig.~\ref{fig:1}(c) for different tunneling rates $\rm t$ and exciton densities $n_{x}$. 
The tunability of the resistivity originates from exciton-hole scattering, as the perpendicular electric field $E_z$ sweeps through the Feshbach resonance and sets the size of the trion Fermi surface. 
For small $\Delta$ the trion Fermi surface is large, and shrinks for larger values of $\Delta$ until it eventually vanishes for $\Delta \gtrsim \mu_h$. 
The contribution of interactions to the hole resistivity is determined 
by the hole many-body scattering rate, which to order $O(g^2)$ reads 
\begin{equation}
 \begin{aligned}
\text{Im}  \Sigma^R_h(\bold p, \epsilon_{h,\bold p}) 
&=  |g|^2 \! \int \!\tfrac{\mathrm{d}^2 \mb k}{(2\pi)^2}
\big (n_F (\epsilon_{t, \mb p +\mb k })+n_B(\omega_{\mb k })\big)\\
&\qquad \times 2\pi\, \delta(\omega_{\mb k}+\epsilon_{h, \mb p}-\epsilon_{t, \mb p+\mb k}) ,
 \label{eq:self_holes}
 \end{aligned}
\end{equation}
with $n_B$ and $n_F$ the Bose and Fermi distributions. 
The conservation of energy in Eq.~\eqref{eq:self_holes} ensures that the trion energy matches the combined energies of the exciton and hole. This
captures the dominant contribution of the resonant 
peak to the scattering amplitude shown in Fig.~\ref{fig:1}(b), and reflects the existence of a metastable Feshbach molecule. 

At low temperatures, Eq.~\eqref{eq:self_holes} is dominated by processes involving small exciton momentum $\mb k$, where the bosonic population $n_B(\omega_{\mb k})$ is largest. Many-body scattering is then maximized when the energy of a trion is resonant with the energy of a hole nearby the Fermi surface. 
In our model this takes place when the two Fermi surfaces are of the same size, which occurs when $\Delta$ is tuned to
\begin{equation}
\Delta_\star = \big (1- m_h/m_t \big )\,\mu_h
    =(2/3)\,\mu_h .
     \label{eq:deltastar}
\end{equation}
We find that tuning the electric field on resonance strongly enhances the hole resistivity by about an order of magnitude, see Fig.~\ref{fig:1}(c).
The background Drude resistivity $\rho_0^h$ is recovered in the limit 
$\Delta \leq 0$, where the phase space volume satisfying the constraint in Eq.~\eqref{eq:self_holes} vanishes, and for sufficiently strong detuning $\Delta\gg 1$ as scattering becomes off-resonant. 

\textbf{Interaction induced drag transport.---} Although excitons are charge neutral and spatially decoupled from the driven layer, they experience drag effects due to exciton-hole scattering. The resulting drag conductivity $\sigma^x$ can be experimentally measured by separately contacting the lower two layers. 
Because it is entirely interaction-driven, the exciton conductivity is highly tunable with the electric field.
Three distinct regimes are identified (Fig.~\ref{fig:2}):

(i) For $\Delta < \Delta_\star$ the hole Fermi surface is smaller than the trion Fermi surface. Thus holes driven out of equilibrium with $p_x>0$ combine mainly with excitons carrying positive momenta $k_x>0$, depleting the exciton distribution for $k_x>0$. This results in an exciton current $J^x_x<0$ that flows in the opposite direction to the hole current. 

(ii) For $\Delta \simeq \Delta_\star$ the exciton drag $\sigma^{x}$ vanishes and changes sign. Dominant scattering results from excitons with small momenta, which carry negligible current. This is in stark contrast with the hole resistivity for which many-body scattering is most dominant in this regime. 

(iii) For $\Delta>\Delta_\star$, the trion Fermi surface is smaller than the hole Fermi surface. Thus holes with $p_x>0$ dominantly combine with excitons with $k_x < 0$, depleating the exciton distribution for $k_x<0$, which
yields an exciton current $J^x_x>0$ flowing in the direction of the hole current.

The exciton drag increases with the hybridization $\rm t$, see Fig.~\ref{fig:2}. Since excitons are charge neutral, their conductivity scales to leading order as $\sigma^{x} \sim g^2 /\rho^h$. Away from resonance and for small tunneling rates $t \ll 1/\tau_0$ the hole resistivity remains close to its background value $\rho^h \approx \rho_0^h+O(g^2)$. Consequently, the exciton conductivity follows $\sigma^{x} \sim g^2/\rho_0 \sim {\rm t}^2 /\rho_0$, as $g \sim \rm t$. Near resonance, however, the hole resistivity is governed by many-body scattering, which also scales as $\rho^h \sim \rm{t}^2$. 
This implies a saturation of $\sigma^{x}$ with increasing tunneling $\rm t$, see Fig.~\ref{fig:2}. 
This intuitive picture, which focuses on the thermally dominant scattering processes, is confirmed by our calculations, which systematically include all such scattering processes, as detailed in the Supplemental Material~\cite{supp}. Our mechanism should be distinguished from exciton drag resulting from polaron formation in monolayer settings~\cite{Cotle2019}. The exciton flow could be accessed experimentally by measuring the charge currents in the bottom ($j_B=-\sigma^t E- \sigma^x E$) and middle ($j_M=2 \sigma^t E+  \sigma^x E$) layers separately, allowing to extract $\sigma^x = - (2j_B+j_M)/E$.

\textbf{Non-monotonous temperature dependence.---}
 \begin{figure}[t]
 \centering
 \includegraphics[width=0.47\textwidth]{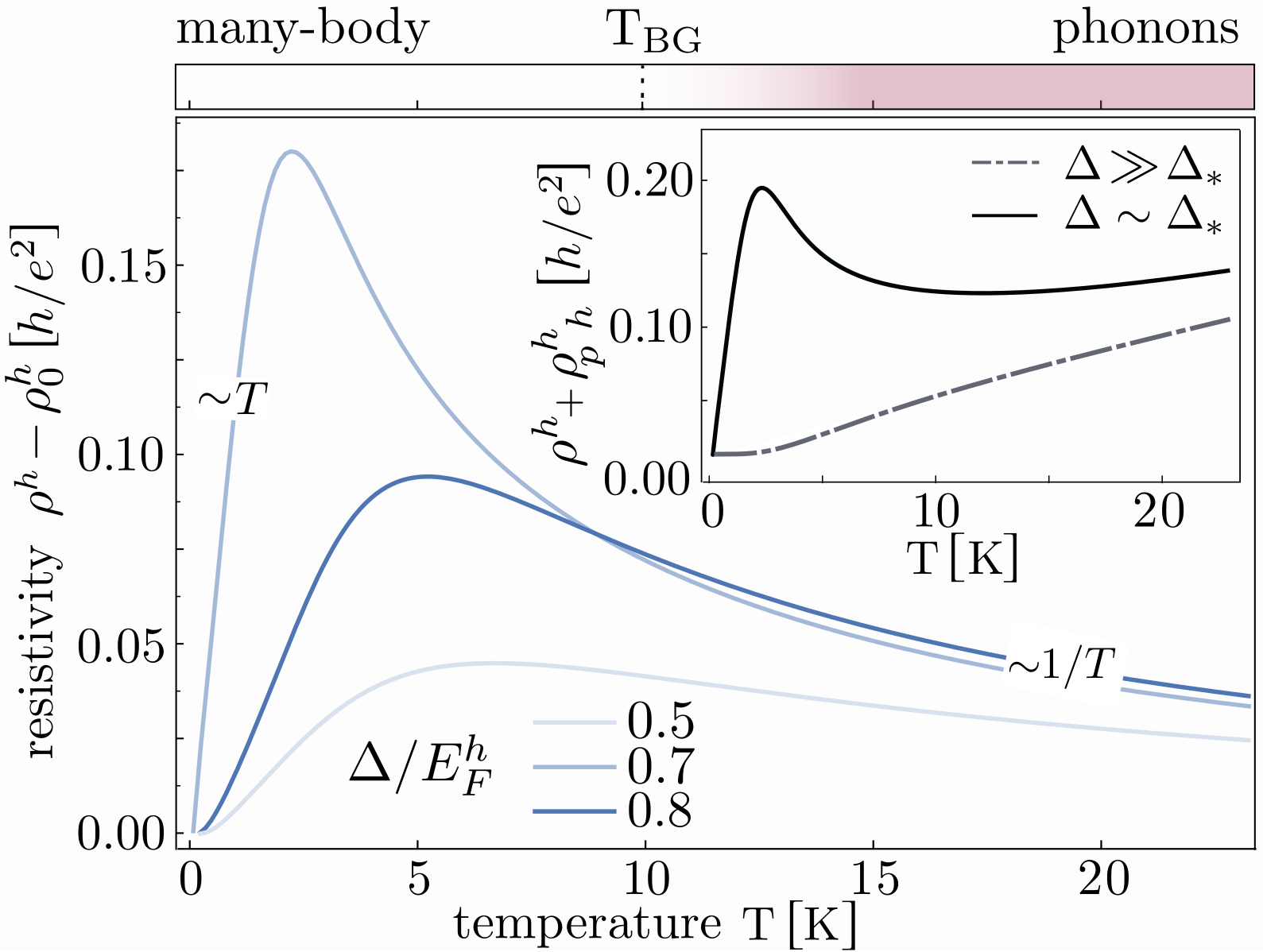}
 \caption{\textbf{Temperature dependence of the hole resistivity.}  Many-body contribution to the resistivity $ \rho^h-\rho^h_{0}$ as a function of temperature $T$ for different detunings $\Delta$, where $\rho^h_{0}=2\pi/\mu_h e^2 \tau_0$ is the Drude resistivity at zero temperature. {Inset:} Total resistivity $\rho^h+ \rho_{\rm ph}^h$ for $\Delta/E_{F}^h = 0.7$, including the contribution from 
scattering with acoustic phonons $\rho^h_{\rm ph}$.
 Below the Bloch-Gr\"uneisen temperature $T_{\rm BG} \approx 10 K$ (consistent with experiments~\cite{Joe2024, Pack2024Jul,Guo2024}), exciton-charge scattering dominates over phonon scattering. 
 }
\label{fig:3c}
 \end{figure}
Conventional scattering due to disorder and phonons typically leads to an electronic resistivity monotonically increasing with temperature $T$. This behavior is in stark contrast to resonant hole-exciton scattering, 
whose strong energy dependence is reflected in the $T$-dependence of resistivity,
see Fig.~\ref{fig:3c}. 
The behavior of the resistivity is divided into three regimes determined by the detuning $\Delta$. The regimes can be understood from the scattering rate $\text{Im} \Sigma^R_h(p_F^h,0)$ in Eq.~\eqref{eq:self_holes} evaluated on the Fermi surface.

(i) At small temperatures $k_B T \ll \hbar^2
n_x/m_x$ and fixed exciton density, the boson chemical potential goes to zero: $\mu_x/k_BT \approx 0$. 
Away from resonance, $(\sqrt{\Delta}-\sqrt{\Delta_\star})^2 \gg k_BT$, we find that $\text{Im} \Sigma^R_h(p_F^h,0)$ decreases exponentially with $1/k_BT$ due to the conservation of energy and momentum.

(ii) At small temperatures  $k_B T \ll \hbar^2
n_x/m_x$ but close to resonance, $(\sqrt{\Delta}-\sqrt{\Delta_\star})^2 \ll k_BT$, excitons with energies $\omega_{\mb k}\ll T$ participate in the scattering.
For these, $n_B(\omega_{\mb k}) \sim T/\omega_{\mb k}$ is large and dominates the momentum integral, yielding a dominant linear-in-$T$  resistivity.

(iii) At high temperatures $k_BT \gg \hbar^2 n_x/m_x$, 
the exciton chemical potential is sizeable, 
and the dominant term in the scattering rate is $\text{Im}\Sigma^R_h(p_F^h, 0) \sim 1/k_BT$.

The many-body contribution dominates the hole resistivity below the Bloch-Grüneisen temperature $T_{\rm BG} = 2\hbar v_s p_F^h$, where $v_s$ is the speed of sound and $p_F^h$ the hole Fermi momentum. At higher temperatures, acoustic phonons with momenta close to $2 p_F^h$ are thermally excited and contribute as $\rho_{\rm ph}^h= 2\pi/\mu_h e^2 \tau_{\rm ph}$ to the resistivity, which increases linearly with $T$. The crossover to phonon-dominated scattering is shown in the inset of Fig.~\ref{fig:3c}, where we have used $\tau_{\rm ph}^h$ estimated in Ref.~\cite{Lavasani2019, Huang2024}.

\begin{center}
\begin{figure}
    \centering    \includegraphics[width=\columnwidth]{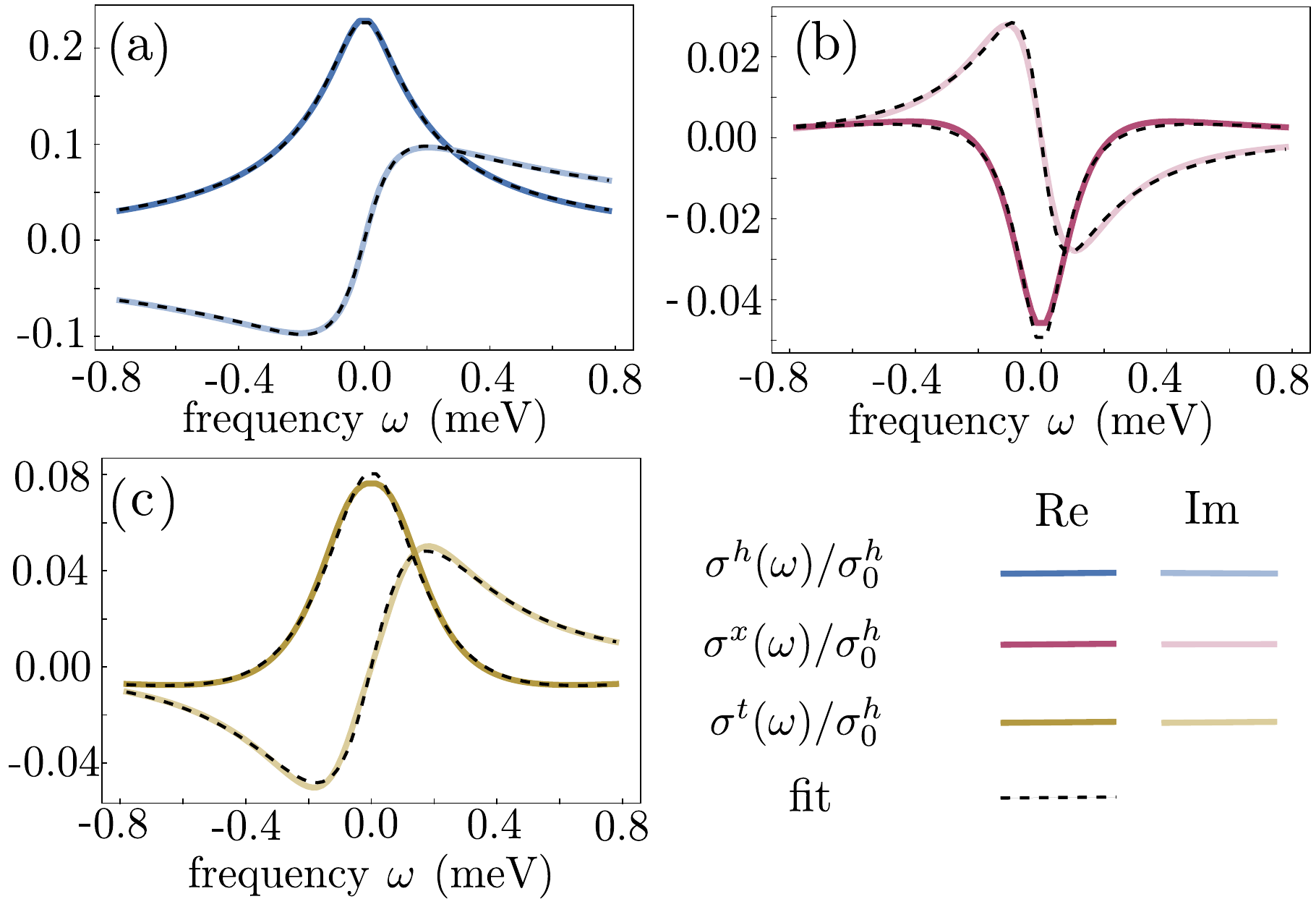}
    \caption{\textbf{Ac conductivities.} 
    Ac conductivity of (a) holes, (b) excitons and (c) trions, at $\Delta/ E_F^h=0.5$ and $T=5.8$ K obtained from the kinetic approach (solid lines) and fitted to the three-fluid model (dashed lines). 
    }
    \label{fig:4}
\end{figure}
\end{center}

\textbf{Ac transport properties.---}
Having analyzed the Feshbach tunable  transport of our Bose-Fermi mixture, we now focus on the ac response, see Fig.~\ref{fig:4}. Albeit challenging to measure experimentally at this point, the ac response provides us with insights in the underlying coupled transport mechanism. We find that the ac response can be effectively described by a simple three-fluid model in terms of the hydrodynamic variables $\bs v_h, \bs v_x, \bs v_t$, which identify as the velocities of the fluids~\cite{footnote_hydrodynamic-variables}.

\begin{subequations}
\begin{align*}
    \dot{\bs v}_h-\frac{e \mb E}{m_h} +\frac{\bs v_h}{\tau_h} &= - \alpha_{th} \frac{n_t}{m_h} \,(\bs v_{h}-\bs v_{t}) -\alpha_{xh} \frac{n_x}{m_h}\, (\bs v_{h}-\bs v_{x}),\\
     \dot{\bs v}_x +\frac{\bs v_x}{\tau_x} &= - \alpha_{tx} \frac{n_t}{m_x}\, (\bs v_{x}-\bs v_{t}) -\alpha_{xh} \frac{n_h}{m_x} \, \bs (\bs v_{x}-\bs v_{h}),\\
      \dot{\bs v}_t + \frac{\bs v_t}{\tau_t} &= - \alpha_{th} \frac{n_h}{m_t}\,(\bs v_{t}-\bs v_{h}) -\alpha_{tx}\frac{n_x}{m_t}\,\bs (\bs v_{t}-\bs v_{x}),
\end{align*}
\end{subequations}
where $\dot{\bs v}_i$ is the time derivative of ${\bs v}_i$.
Three drag coefficients $\alpha_{ij}$, with $i\neq j \in \{h,x,t\}$, 
model the effect of interactions in the Bose-Fermi mixture as ``viscous friction forces." 
The three momentum relaxation rates $\tau_i$, for $i\in \{h,x,t\}$, model the many-body corrections to $\tau_0$. Only holes feel the longitudinal acceleration field $e \mb E / m_h$.
The right hand side of the equations is the most general system of linear coupling terms which vanishes at $\bs v_h=\bs v_x=\bs v_t$ and preserves the total momentum $\bs P = \sum_{i} n_i m_i \bs v_i$ (i.e.\ they ensure $\dot{\bs P}=\bs 0$ in the limit $\tau_i \rightarrow \infty$ and at $\mb E = \bs 0$).
Solving these equations, we obtain the ac conductivities  from $n_i \bs v_i = \sigma^i\, \mb E/e$ as functions of six parameters $\tau_h,\tau_x,\tau_t,\alpha_{xh},\alpha_{th},\alpha_{tx}$ which are  functions of $\Delta$ and $T$; for instance, the sign-switching exciton drag coincides with a sign-switching drag coefficient $\alpha_{xh}\propto \text{sign}(\Delta-\Delta_\star)$. 
 A striking feature of the ac results is that the dissipative part of drag conductivities of excitons and trions ${\rm Re}(\sigma^x), {\rm Re}
(\sigma^t)$ change their sign at a finite frequency set by the competition between drag forces and relaxation rates, highlighting the deviations from the Lorentz-shaped Drude behavior.

\textbf{Outlook.---} In this work we study tunable transport properties of a strongly-correlated Bose-Fermi mixtures realized in TMD heterostructures. As a result of the strong interactions near the solid-state Feshbach resonance, we predict an unconventional temperature dependence of the conductivities and a sign-changing exciton drag. Remarkably, selectively enhanced many-body scattering dominates over other scattering channels such as disorder and phonons, provided the system is below the Bloch-Gr\"uneisen temperature. Our results can guide future experiments and demonstrate the potential for TMD heterostructures to investigate strongly interacting Bose-Fermi mixtures, which appear in a variety of physical settings and often undergo interesting instabilities.

%In our analysis, we did not account for the effects of %such as Wigner crystallization of the holes or condensation instabilities of the excitons, leaving these aspects for future investigation. A Wigner crystal phase can form only at low doped charge densities. Since our results do not qualitatively depend on the specific densities considered, this regime can be easily avoided experimentally. Nevertheless, as such phases can occur in these multi-layer structures, future studies of their impact on transport properties could be highly intriguing. Regarding
%Notably, the present work does not address the possible exciton superfluid phases. 
Our findings open exciting avenues for exploring the interplay between transport phenomena and pairing instabilities in Bose-Fermi mixtures allowed by exciton condensation. In fact, the low relative temperatures ($T/T_F \leq 0.01$) already achieved experimentally~\cite{Pack2024Jul,Joe2024,Guo2024,Nguyen23} are promising for reaching regimes of exciton condensation. Phase fluctuations in the exciton gas are expected to play a significant role~\cite{Ma2021,Cotlet16}, potentially leading to the emergence of unconventional phases~\cite{Zerba2024,kumar2024}.
Furthermore, additional theoretical studies could provide insights into the hydrodynamic behavior of these quantum mixtures~\cite{Levchenko2020,Lucas2021,Fritz2024}. Another interesting direction for future work is to explore how the stability of Wigner crystals, realized in the low density regime of holes~\cite{Smoleski2021,Zhou2021}, is modified by the presence of a finite exciton density.

\textit{\textbf{Acknowledgments.---}} We thank Z. Hao, A. Imamo\u{g}lu, W. Kadow, and A. Mozes for fruitful discussions. We acknowledge support from the Deutsche Forschungsgemeinschaft (DFG, German Research Foundation) under Germany’s Excellence Strategy--EXC--2111--390814868, TRR 360 – 492547816 and DFG grants No. KN1254/1-2, KN1254/2-1, the European Research Council (ERC) under the European Union’s Horizon 2020 research and innovation programme (grant agreement No. 851161), as well as the Munich Quantum Valley, which is supported by the Bavarian state government with funds from the Hightech Agenda Bayern Plus. C.K. acknowledges funding from the Swiss National Science Foundation (Postdoc.Mobility Grant No. 217884).

\textit{\textbf{Data availability.---}} Data and codes are available upon reasonable request on Zenodo~\cite{zenodo}.

\textit{\textbf{Author contributions.---}} C.Z. and L.M. developed the kinetic theory. C.K. developed the Kubo formalism. C.Z. performed the numerical analysis. M.K. conceived the project. All authors contributed to the discussions of the results and the writing of the manuscript.

\bibliography{library}
\clearpage

\appendix
\section{Microscopic model} 
\begin{figure*}
    \centering
    \includegraphics[width=0.89\linewidth]{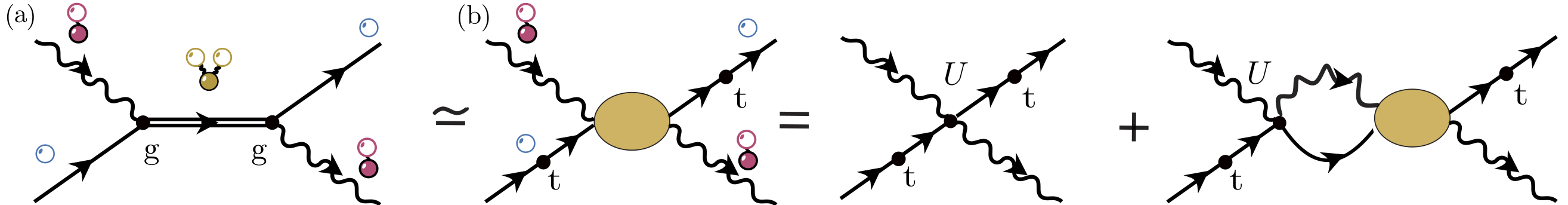}
    \caption{\textbf{Diagrammatic derivation of the effective model.} \textbf{a)} T-matrix $T_{\rm eff}$ for the effective interaction between hole, exciton and trion. \textbf{b)} The T-matrix $T$ of the microscopic model is given by the solution of the Bethe-Salpeter equation; at leading order and near the Feshbach resonance, it is equivalent to the T-matrix $T_\text{eff}$ of the effective model.}
    \label{fig:diag}
\end{figure*}
Here we derive the effective model of the main text from a microscopic Hamiltonian which describes the scattering between holes and excitons across different layers:
\begin{align}
\hat{H} =& \sum_{\mathbf{k}} x^\dagger_{\mathbf{k}} \frac{k^2}{2m_x} x_{\mathbf{k}} +\begin{pmatrix}c^\dagger_{\mathbf{k},T} \\ 
c^\dagger_{{\mathbf{k}},M}\end{pmatrix} \begin{pmatrix} \frac{{k}^2}{2m}+ \Delta & \rm t \\ \rm t & \frac{k^2}{2m}\end{pmatrix}\begin{pmatrix} c_{{\mathbf{k}},T} \\ 
c_{{\mathbf{k}},M}\end{pmatrix}\nonumber\\
&+ \frac{U}{V} \sum_{\mathbf{k},\mathbf{k}',\mathbf{q}} c_{{\mathbf{k}},M}^\dagger c_{{\mathbf{k}}+{\mathbf{q}},M} x^\dagger_{{\mathbf{k}'}} x_{{\mathbf{k}'}-{\mathbf{q}}},
\label{Eq:BoseFermiHamiltonian}
\end{align}
where the indices $\{ T, M \}$ label the top and middle layers, $\rm t$ is the amplitude for doped holes tunneling from the top layer to the middle layer, and we assume contact interactions of strength $U$ between interlayer excitons and middle-layer holes.

Scattering between excitons and holes is described by going to the center-of-mass frame, in which the wavefunction at energy $E=\mb k^2/2m_{\rm red}$ takes the following form
\begin{equation}
    \psi_\mathbf{k}(\mathbf{r}) = e^{i\mathbf{k}\mathbf{r}} + f(\mathbf{k}) \sqrt{\frac{i}{8\pi}}\frac {e^{i|\mathbf{k}|\mathbf{r}|}}{\sqrt{|\mathbf{r}||\mathbf{k}|}},
    \label{eq:scattering}
\end{equation}
where $\mathbf{k}$ denotes the relative momentum, $e^{i\mathbf{k}\mathbf{r}} $ is an incoming plane wave and the last term describes the outgoing spherical wave after scattering. Eq.~\eqref{eq:scattering} defines a dimensionless scattering amplitude $f(\mathbf{k})$, which is related to the cross-section $d\sigma/d\theta$ via
\begin{equation}
    \frac{d\sigma}{d\theta}(\mb k) = \frac{|f(\mathbf{k})|^2}{8\pi|\mathbf{k}|}.
\end{equation}
One can also relate $f(\mb k)$ to the T-matrix, by analyzing the Lippmann-Schwinger equation for the scattering state $|\psi_\mathbf{k}^+\rangle$ (identified with $\psi_\mathbf{k}(\mathbf{r})$ in Eq.~\eqref{eq:scattering}):
\begin{equation}
  |\psi_\mathbf{k}^+\rangle =   |\mathbf{k}\rangle  + G_0 \hat H_{xh} |\psi_\mathbf{k}^+\rangle = |\mathbf{k}\rangle  + G_0 \hat{T} |\mathbf{k}\rangle,
  \label{eq:LipSch}
\end{equation}
where $\hat H_{xh}$ is the interaction term corresponding to the second line of Eq.~\eqref{Eq:BoseFermiHamiltonian} and $\langle \mathbf{r} |G_0 | \mathbf{r}'\rangle = \int \frac{
d^2k}{(2\pi)^2}  \frac{e^{i \mathbf{k}(\mathbf{r}-\mathbf{r}')}}{E +i0^+ - \mathbf
k^2/2m_{\rm red}}$. Expanding Eq.~\eqref{eq:LipSch} for large $|\mathbf{r}|$ establishes the relation between $\hat{T}$ and $f(\mathbf{k})$:
\begin{equation}
    f(\mathbf{k})= -2m_{\rm red}\, T(\mathbf{k}^2/2m_{\rm red}).
\end{equation}
This defines the dimensionless $f(E)=f(\mb k)\big |_{\mathbf{k}^2/2m_{\rm red}=E}$ in Fig.~\ref{fig:1} of the main text.

Starting from Eq.\eqref{Eq:BoseFermiHamiltonian}, the hole-exciton T-matrix is obtained by solving the Bethe-Selpeter equation depicted in Fig.~\ref{fig:diag} (b), and reads $T=\frac{\rm t^2}{U^{-1}-\Xi_{xh}}G^2_{h,M}$, where $G_{h,M}$ is the propagator of holes in the middle layer and $\Xi_{xh}$ is the exciton-hole bubble. Since $\Xi_{xh}$ is UV divergent, we regularize the theory by introducing a momentum cutoff and renormalizing $U$ such that the model reproduces the experimental value of the inter-layer trion binding energy $E_t^0$ (see Ref.~\cite{Clemens21}). Near a Feshbach resonance, the scattering process is predominantly governed by this interlayer trion, %Using a renormalization procedure to obtain $U_0$ as a function of $U$ and $E_t^0$ the experimental bound state energy (see Ref.~\cite{Clemens21})
 which yields the following T-matrix near resonance:
\begin{subequations}
\begin{align}
 T(\omega) &\simeq-\frac{2\pi}{m_\text{red}\Delta^2\log{((\omega-\Delta)/E^{0}_{t}})} {\rm t^2} \\
 & \simeq \frac{2\pi| E^0_t|}{m_\text{red} \Delta^2 (\omega - E^0_t + \Delta + i0^+)} {\rm t^2}.
\end{align}
\end{subequations}
 The microscopic T-matrix near resonance is found to have a pole at the energy of the molecule. 
Consequently, we can consider an effective model where the scattering between excitons and top-layer electrons is mediated by a virtual trion. The elementary scattering event of the effective model, represented by the diagram in Fig.~\ref{fig:diag} (a), is described by an effective T-matrix  $T_{\rm eff}$, which has a pole at the trion energy:
\begin{equation}
 T_{\rm eff}(\omega) \approx \frac{g^2}{\omega - E^0_t+ \Delta + i0^+} ,
\end{equation}
see also Refs. \cite{Stoof2009,Zwerger2016,Zerba2024}. The coupling parameter $g$ is the effective hole-exciton-molecule vertex. By matching the effective T-matrix \( T_{\rm eff} \) to second order in \( g \) with the full T-matrix \( T \), we can derive an expression for \( g \),
\begin{equation}
g = \left({\frac{2\pi}{m_\text{red} E_{t}^0 }}\right)^{\frac{1}{2}} \rm t,
\end{equation}
which is the effective three-body coupling of \eq{eq:eff_model}. 

%For a two-dimensional system, the regime of strong coupling is not reached in the limit of a diverging scattering length, as the scattering amplitude takes the universal form $f(k) \sim \log(k\,a_{2D})^{-1}$, and  vanishes for small momenta $k\rightarrow 0$ irrespective of the scattering length $a_{2D}$. Instead, the regime of a strongly enhanced scattering cross section is achieved when the molecule has entered the continuum and has become a meta-stable resonant excitation; see Fig.~1(b) of the main text. 

\section{Kinetic equations}
\label{App:A}

\textbf{Derivation.---} Starting from the Hamiltonian Eq.~\eqref{eq:eff_model}, we derive a set of three coupled Boltzmann equations,
\begin{align}
  \label{eq:1}
  \Big (   \partial_t + \bold v_{i} \cdot \bs \partial_{\mb r} + \mb f_{i} \cdot \bs \partial_{\mb q} \Big ) F_i &= I_i [F_h, F_x, F_t],
\end{align}
for the three species of particles, $i\in\{h,x,t\}$, holes, excitons, trions, respectively. These equations provide a semiclassical approximation of the dynamics of the system out of equilibrium, where $F_i(\bold r , \bold q,t)$ are the quasiparticle mass-shell distribution functions in phase space (at position $\mb r$ and momentum $\mb q$), to be determined.
On the left-hand side, $\mb f_i(\mb r,\mb q)$ is the force acting on a particle and $\mb v_i(\mb r,\mb q)$ is the particle's velocity. 
The right-hand side $I_i$ is the collision integral, which accounts for the effect of out-of-equilibrium many-body scattering as well as incoherent background scattering arising for example from impurities.

We use the real-time formalism for nonequilibrium field theory \cite{Kamenev2011Sep,Rammer} to obtain the hole-exciton-trion contribution to the kinetic equations Eq. \eqref{eq:1}.
This involves defining the quantum action along a closed time contour going forward from $t=-\infty$ to $t=+\infty$, then backward from $t=+\infty$ to $t=-\infty$, and for each a different set of quantum fields. In coherent state notations where $\bar x,\bar h,\bar t$ represent the conjugate fields of $x,h,t$ respectively, we thus introduce $x^\pm,\bar x^\pm,h^\pm,\bar h^\pm,t^\pm,\bar t^\pm$ where $+$ labels operators appearing in the forward-time integral, and  $-$ in the backward-time integral.
We recall the Keldysh rotation for a bosonic field $x$
\begin{subequations}
\begin{align}
  \label{eq:2}
  x_{\rm cl/q} (t)&= \big ( x^+(t) \pm x^-(t) \big ) /\sqrt 2,\\
 \bar x_{\rm cl/q} (t) &= \big ( \bar x^+(t) \pm \bar x^-(t) \big ) /\sqrt 2,
\end{align}
\end{subequations}
where $+$ is for ``cl" and $-$ for ``q" in $\pm$, 
while for a fermionic field $f$ (here $f=h,t$) 
\begin{subequations}
  \label{eq:3}
  \begin{align}
  f_{\rm r/a} (t) &= \big ( f^+(t) \pm f^-(t) \big ) /\sqrt 2 ,\\
  \bar f_{\rm r/a} (t) &= \big ( \bar f^+(t) \mp \bar f^-(t) \big ) /\sqrt 2 ,
  \end{align}
\end{subequations}
where $+$ is for ``r" and $-$ for ``a" in the first line and conversely in the second.
The action for the free fields ($f=h,t$) reads 
\begin{align}
  \label{eq:5a}
  S_{x,0} &=  \sum_{\mb k,\mb k'} \int \text
    dt \begin{bmatrix}
      \overline x^{\rm cl}\\
     \overline x^{\rm q}
    \end{bmatrix}^\top_{\mb k} 
    \begin{bmatrix}
      0 & [G_{x,0}^{-1}]^{\rm A}\\
  [G_{x,0}^{-1}]^{\rm R}& [G_{x,0}^{-1}]^{\rm K}
    \end{bmatrix}_{\mb k,\mb k'}
    \begin{bmatrix}
      x^{\rm cl}\\
      x^{\rm q}
    \end{bmatrix}_{\mb k'}, \\
      \label{eq:5b}
   S_{f,0} &=   \sum_{\mb p,\mb p'} \int \text
    dt  \begin{bmatrix}
      \overline f^{\rm r}\\
      \overline f^{\rm a}
    \end{bmatrix}^\top _{\mb p}
    \begin{bmatrix}
    [G_{b,0}^{-1}]^{\rm R}   & [G_{b,0}^{-1}]^{\rm K}\\
  0 & [G_{b,0}^{-1}]^{\rm A}
    \end{bmatrix}_{\mb p,\mb p'}
    \begin{bmatrix}
      f^{\rm r}\\
      f^{\rm a}
    \end{bmatrix}_{\mb p'}.
\end{align}
Here and in the following, time integrals are defined along the forward contour. 

Eqs.~\eqref{eq:5a},\eqref{eq:5b} involve the free (retarded and advanced) bosonic propagator $[G_{x,0}^{\rm R/A}]_{\mb k,\mb k'}(E)= \frac{\delta_{\mb k,\mb k'}}{E-\omega_{\mb k}\pm i0^+}$
and fermionic propagator $[G_{f,0}^{\rm R/A}]_{\mb p,\mb p'}(E) = \frac{\delta_{\mb p,\mb p'}}{E-\epsilon_{f,\mb p}\pm i0^+}$, where $+$ is for ``R" and $-$ is for ``A" in $\pm$.

The free Keldysh propagators are for bosons $[G_{x,0}^{\rm K}]_{\mb k,\mb k'}(E) = -2i\pi F_x(\mb k,E)\delta(E-\omega_{\mb k})\delta_{\mb k,\mb k'}$ and for fermions $[G_{f,0}^{\rm K}]_{\mb p,\mb p'}(E)  = -2i\pi F_f(\mb p,E)\delta(E-\epsilon_{f,\mb p})\delta_{\mb p,\mb p'}$; at this stage $F_i(\mb q,E)$ are just parameter functions. 

The interacting part of the action, in terms of the Keldysh-rotated fields, reads
\begin{align}
  \label{eq:7}
  S_{\rm int} &= - \frac g {\sqrt{2 N}} \int \text dt \sum_{\mb k,\mb p} 
                \Big (
                x_{\mb k}^{\rm cl} \,\bar t_{\mb p+\mb k}^{\rm a}\, h_{\mb p}^{\rm a}
                + x_{\mb k}^{\rm cl} \,\bar t_{\mb p+\mb k}^{\rm r} \, h_{\mb p}^{\rm r}\\
                &
                \qquad \qquad + x_{\mb k}^{\rm q}\, \bar t_{\mb p+\mb k}^{\rm a}\, h_{\mb p}^{\rm r} 
                  +  x_{\mb k}^{\rm q} \,\bar t_{\mb p+\mb k}^{\rm r}\, h_{\mb p}^{\rm a}  \Big ) +\rm h.c. \nonumber ,
\end{align}
where ``+ h.c.'' means conjugating all fields (in the coherent state path integral sense), exchanging r/a and commuting the fermionic Grassmann fields. Here $N$ is the total number of unit cells in a layer, and in the following $A$ will be the total area of a layer.

From Eq. \eqref{eq:7}, the self-energies can be obtained perturbatively from Dyson's equation, 
\begin{align}
  \label{eq:dyson}
\bs G_i(t,t')&= \bs G_{0,i}(t-t') \\
 &+\iint \text dt_1 \text dt_2 \bs G_{0,i}(t-t_1)\bs \Sigma_i(t_1,t_2) \bs G_{i}(t_2-t')\nonumber,
\end{align}
for $i\in {h,x,t}$ and 
where all quantities $\bs G_{i},\bs G_{0,i},\bs \Sigma_i$ have a matrix structure in momentum space (i.e. they are functions of two momentum variables)
and in Keldysh space (they have a $2\times 2$ structure with indices $\{\text{cl,q}\}$ for the bosons and $\{\text{r,a}\}$ for the fermions). 
The interacting theory retains the same matrix structure as the free theory, thus defining retarded, advanced  and Keldysh self energies~\cite{Kamenev2011Sep,Rammer}; for bosons  $\Sigma_x^{\rm R}= \Sigma_x^{\rm q,cl}\, ,\,\,
  \Sigma_x^{\rm A}= \Sigma_x^{\rm cl,q}  \, ,\,\,
   \Sigma_x^{\rm K}= \Sigma_x^{\rm q,q} $, and for fermions $\Sigma_{f}^{\rm R}= \Sigma_{f}^{\rm r,r} \, ,\,\,
  \Sigma_{f}^{\rm A}= \Sigma_{f}^{\rm a,a} \, ,\,\,
   \Sigma_{f}^{\rm K}= \Sigma_{f}^{\rm r,a} $ where $f\in\{h,t\}$.
   
The kinetic theory is obtained as a semiclassical approximation of the time evolution equations in phase space. This is obtained by computing the Wigner transform of the self-energy, 
\begin{align}
 \label{eq:wigner}
 \bs \Sigma_i[\mb R,t;\mb k, E] &= \textstyle{\int \text dt' \, e^{i E  t'}  \sum_{\mb k'} } e^{i\mb k' \mb R}\nonumber \\
    &\times \bs \Sigma_i(\mb k+\tfrac{\mb k'}2,\mb k- \tfrac{\mb k'}2, t+\tfrac {t'}2, t-\tfrac {t'}2),
\end{align}
where $\mb R,t$ are the ``slow" space and time variables, and $\mb k,E$ are the ``fast" momentum and energy variables.
To the perturbative order $O(g^2)$, we find
\begin{subequations}
\begin{align}
\label{eq:sigmaAR}
  \Sigma^{\rm R/A}_x(\mb k, E)
  &= \frac {g^2}{2A}\sum_{\mb p}  \frac {  F_t(\mb p+\mb k) -    F_h(\mb p)  }{  E - \epsilon_{t,\mb p+\mb k} + \epsilon_{h,\mb p} \pm i 0 },\\
  \Sigma^{\rm R/A}_h( \mb p, E ) 
  &= -\frac{g^2}{2A}\sum_{\mb k}\frac{F_t( \mb k+\mb p)-F_x(\mb k)}{E+ \omega _{\mb k}-\epsilon_{t, \mb k+\mb p}\pm i0},\\
    \Sigma^{\rm R/A}_t(\mb p, E )
 &=\frac{g^2}{2A}\sum_{\mb k} \frac{F_x(\mb k)+F_h(\mb p -\mb k)}{E-\omega_{\mb k }- \epsilon_{h,\mb p -\mb k} \pm i0},
\end{align}
\end{subequations}
where $+$ is for R and $-$ is for A in $\pm$, and 
 \begin{subequations}
    \begin{align}
   \Sigma^{\rm K}_x(\mb k, E)
    &= -\frac {i\pi g^2}{A}\sum_{\mb p}\Big ( 1 -F_t(\mb p+\mb k) F_h(\mb p)  \Big )\nonumber \times\\&\times
    \delta \Big ( E  -\epsilon_{t,\mb p+\mb k} +\epsilon_{h,\mb p} \Big ),\\
    \Sigma^{\rm K}_h(\mb p, E )
    &= \frac{i\pi g^2}{A}\sum_{\mb k}\big( 1-F_t(\mb k+\mb p)F_x(\mb k)\big) 
       \times\nonumber\\&\times\delta(E+ \omega _{\mb k}-\epsilon_{t, \mb k+\mb p}) ,\\
  \Sigma^{\rm K}_t(\mb p, E ) 
  &= - \frac{i\pi g^2}{A}\sum_{\mb k} (F_x(\mb k)F_h(\mb p - \mb k)+1)
     \times \nonumber\\&\times\delta(E-\omega_{\mb k }- \epsilon_{\mb p -\mb k})   .
     \label{eq:sigmaK}
\end{align}
 \end{subequations}

 In Eqs.~\eqref{eq:sigmaAR}--\eqref{eq:sigmaK} and all the following, we keep implicit the dependence of distributions $F_i$ on the slow variables $\mb R,t$. 
 The collision integral is obtained from the the Wigner transform of the self-energy as
\begin{equation}
\label{eq:IfromSigma}
\tilde I_i(\mb q) = i \Sigma^{\rm K}_i(\bold q, \xi_{i,\mb q}) + 2 {\rm Im}\Sigma^{\rm R}_i(\bold q, \xi_{i,\mb q}) F_i(\mb q,\xi_{i,\mb q}),
\end{equation}
with the shorthand $\xi_{x,\mb q}=\omega_{\mb q}$ and $\xi_{f,\mb q}=\epsilon_{f,\mb q}$. 
In general $F_i(\bold q ,E)$ is a function of energy and momentum independently, however in Eq.~\eqref{eq:IfromSigma} 
the energy variable is locked to the particle dispersion: this is a semiclassical approximation which relies on the existence of well-defined quasiparticles and allows one to write simply $F_i(\mb q)$ where the on-shell energy argument $\xi_{i,\mb q}$ is implicit.

Inserting Eqs.~\eqref{eq:sigmaAR}--\eqref{eq:sigmaK} into Eq.~\eqref{eq:IfromSigma} gives
\begin{widetext}
    \begin{subequations}
    \begin{align}
 \tilde I_x &= \frac{g^2  \pi}{A} \sum_{\mb p}    \delta \Big ( \omega_{\mb k}  -\epsilon_{t,\mb p+\mb k} +\epsilon_{h,\mb p} \Big )
       \bigg [ \big ( 1 -F_t(\mb p+\mb k) \; F_h(\mb p)  \big ) 
       +\big (  F_h(\mb p) -  F_t(\mb p+\mb k) \big )  F_x(\mb k)  \bigg ] ,\\
\tilde I_h &=- \frac{g^2  \pi}{ A} \sum_{\mb k}    \delta \Big ( \omega_{\mb k}  -\epsilon_{t,\mb k+\mb p} +\epsilon_{h,\mb p} \Big )
        \bigg [ \big ( 1 -F_t(\mb k+\mb p)\; F_x(\mb k)  \big )
        +\big (  F_x(\mb k) -  F_t(\mb k+\mb p) \big )  F_h(\mb p)  \bigg ] ,\\
\tilde I_t &= \frac{g^2  \pi}{A}\sum_{\mb k}  \delta \Big ( \omega_{\mb k}  +\epsilon_{h,\mb p - \mb k} -\epsilon_{t,\mb p} \Big )
        \bigg [ \big ( 1 +F_x(\mb k)\; F_h(\mb p - \mb k)  \big ) 
        -\big (  F_x(\mb k) +  F_h(\mb p - \mb k) \big )  F_t(\mb p)  \bigg ].
\end{align}
\end{subequations}
\end{widetext}

These collision integrals satisfy detailed balance: they vanish when $F_i$ are replaced by their equilibrium values
$F_x^{\rm eq}(\mb k) = 1+ 2 n_{\rm B}(\omega_{\mb k})$ and $F_h^{\rm eq}(\mb p) = 1 - 2 n_{\rm F}(\epsilon_{h,\mb p}) $ and $F_t^{\rm eq}(\mb p) = 1 - 2 n_{\rm F}(\epsilon_{t,\mb p}) $.
Thus $\tilde I_i$ capture interaction effects out of equilibrium, however they preserve total momentum \cite{baym1961conservation} and do not account for current relaxation. In other words, $\tilde I_i$ may capture drag effects but not diffusive transport. We then include momentum relaxation for all three species in the relaxation time approximation, at the rates $\tau_{h,\mb p}, \tau_{x,\mb k}, \tau_{t,\mb p}$. Such a collision term can arise from incoherent background scattering, for instance due to impurities.
Thus, in Eq.~\eqref{eq:1} we take $I_i[F_h,F_x,F_t] = \tilde I_i[F_h,F_x,F_t] - (F_i-F_i^{\rm eq})/\tau_i $.

Besides, in Eq.~\eqref{eq:1} all the energies, velocities and quasiparticle weights are in principle renormalized by interactions. However, such corrections are always higher order in powers of $g^2$ or $\tau_0/E_{\rm F}^h$ than our level of approximation, thus we neglect such corrections and use the bare dispersions $\omega_{\mb k}, \epsilon_{h,\mb p}, \epsilon_{t,\mb p}$ and velocities $\bold v_{x}(\mb k)=\partial_{\bold k} \omega (\bold k)$, $\bold v_{h}(\mb p)=\partial_{\bold p} \epsilon_h (\bold p)$, $\bold v_{t}(\mb k)= \partial_{\mb p} \epsilon_t (\bold p)$. 
 In the setup we consider, the electric field $\mb E$ applies solely to the holes, thus $\mb f_{h, \mb p} = e \mb E$  and  $\mb f_{x, \mb k}=0= \mb f_{t, \mb p} =0$. 
We furthermore assume a homogenous system (where temperature and chemical potentials are uniform), so that all gradient terms vanish in equilibrium, $\mb v_i \bs \partial_{\mb r}F_i^{\rm eq}= 0$. 

\textbf{Solution.---} We proceed by linearizing the kinetic equations Eq.~\eqref{eq:1} around equilibrium. We parameterize the non-equilibrium distributions as $F_x(\mb k) = F_x^{\rm eq}(\mb k)+ 2 \varphi_x(\mb k)$ and $F_h(\mb p) = F_h^{\rm eq}(\mb p) - 2 \varphi_h(\mb p)$ and $F_t(\mb p) = F_t^{\rm eq}(\mb p) - 2 \varphi_t(\mb p)$ for excitons, holes and trions respectively, $n_{\rm B}, n_{\rm F}$ are the Bose and Fermi functions, and $\varphi_h,\varphi_{x},\varphi_t$ are the deviations from equilibrium of the particle populations. 
From these, the conductivities are then defined as 
\begin{equation}
\label{eq:conductivitiesfromphi}
  \sigma^i(\Omega) \mb E = e\int \frac{\text{d}^2\bold{k}}{(2\pi)^2} \, \,\bold v_{i}(\mb k)\, \varphi_i(\bold k )
\end{equation}
for $i \in {h, x, t}$.
The linear response of the system to the external electric field is obtained by performing this replacement in the collision integral, while simply replacing $F_i$ by $F_i^{\rm eq}$ on the left-hand side, except for the partial time derivative term which we write, in ac notations, as $i\Omega(F_i-F^{\rm eq}_i)$. As a result, we obtain the set of three coupled algebraic equations
\begin{widetext}
\begin{subequations}
\begin{align}
&\bigg( \frac 1 { \tau_{x,\mb k}} + i\Omega\bigg)  \varphi_x(\mb k) 
= \frac{ 2\pi} {A} g^2\sum_{\mb p}  \delta \Big ( \omega_{\mb k}  -\epsilon_{t,\mb p+\mb k} +\epsilon_{h,\mb p} \Big )
    \bigg [ \big (  n_{\rm F}(\epsilon_{t,\mb p+\mb k}) -  n_{\rm F}(\epsilon_{h,\mb p}) \big )  \varphi_x(\mb k)\nonumber \\
 &  \qquad \qquad + \big (  n_{\rm B}(\omega_{\mb k})  - n_{\rm F}(\epsilon_{h,\mb p})  + 1\big ) \varphi_t(\mb p+\mb k) 
    - \big ( n_{\rm F}(\epsilon_{t,\mb p+\mb k}) + n_{\rm B}(\omega_{\mb k})  \big ) \varphi_h(\mb p) \bigg ],  \label{eq:17} \\
&  \bigg( \frac 1 {   \tau_{h,\mb k}} + i\Omega\bigg) \varphi_h(\mb p )
 =  \frac{2\pi}{A} g^2 \sum_{\mb k }\delta\big(\omega_{\mb k}+\epsilon_{h, \mb p }- \epsilon_{t, \mb p+ \mb k }\big)
    \bigg[ \varphi_{x}(\mb k)\big( n_{\rm F} (\epsilon_{t,\mb k + \mb p })- n_{\rm F} (\epsilon_{h, \mb p })  \big)\nonumber \\
  & \qquad \qquad +\varphi_{t}(\mb k+\mb p) \big( n_{\rm B}  (\omega_{\mb k}) - n_{\rm F}  (\epsilon_{h, \mb p }) + 1 \big) -\varphi_{h}(\mb p )
    \big( n_{\rm B}  (\omega_{\mb k})  + n_{\rm F} (\epsilon_{t, \mb p+ \mb k })\big)  \bigg ]
  - \Big ( \bs f_{h,\mb p} \cdot \bs \partial_{\mb p} \Big )   n_{\rm F}(\epsilon_{h,\mb p}) , \label{eq:18} \\
&\bigg( \frac 1 { \tau_{t,\mb p}} + i\Omega\bigg)\varphi_{t}(\mb p)  
= \frac{2\pi}{A} g^2\sum_{\mb k }\delta\big(\omega_{\mb k}+\epsilon_{h, \mb p- \mb k }- \epsilon_{t, \mb p }\big)
    \bigg[ \varphi_{x}(\mb k)\big( n_{\rm F} (\epsilon_{h,\mb p -\mb k}) - n_{\rm F} (\epsilon_{t, \mb p}) \big)\nonumber \\
  &\qquad \qquad -\varphi_{t}(\mb p )\big(n_{\rm B} (\omega_{\mb k}) - n_{\rm F} (\epsilon_{h,\mb p -\mb k}) + 1\big)  + \varphi_{h}(\mb p-\mb k)
    \big( n_{\rm B} (\omega_{\mb k}) + n_{\rm F} (\epsilon_{t, \mb p}) \big) \bigg] . \label{eq:19}  
\end{align}
\label{eq:eqnsforvarphi}
\end{subequations}

In what follows, we use matrix notations for objects that are function of two momenta. In particular, we define the vector $ \mathcal K^h_{\mb p}=  \Big ( \mb f_{h,\mb p} \cdot \bs \partial_{\mb p} \Big ) \,n_{\rm F}(\epsilon_{h, \mb p})$ and the diagonal matrices
$\Upsilon^i_{\mb q,\mb q'} =\delta_{\mb q, \mb q'} \bigg(\frac{1} { \tau_{i,\mb q}}+ i\Omega\bigg)$ and
\begin{equation}
    \mathcal M^t_{\bold p, \bold q}
= \delta_{\bold q, \bold p}     \frac {2\pi}{A}g^2\sum_{\mb k}   \delta \big ( \omega_{\mb k}  +\epsilon_{h,\mb p-\mb k} -\epsilon_{t,\mb p} \big )
\; \Big (  n_{\rm B}(\omega_{\mb k})+1-  n_{\rm F}(\epsilon_{h,\mb p-\mb k}) \Big ).
\end{equation}
We first solve for $\varphi_t(\mb p)$ in terms of 
$\varphi_x(\mb k), \varphi_h(\mb p)$:
\begin{align}
\label{eq:varphit}
    \varphi_t(\bold p) &= \frac {2\pi}{ A}g^2 \sum_{\mb k} \left (\mathcal{M}^t+\Upsilon_t \right )^{-1}_{\bold p, \bold p}
    \delta \Big ( \omega_{\mb k}  +\epsilon_{h,\mb p-\mb k} -\epsilon_{t,\mb p} \Big )\nonumber \\
    &\qquad \qquad \times \bigg ( 
(n_F(\epsilon_{h,\bold{p}-\bold k })-n_F(\epsilon_{t,\bold{p} }))\, \varphi_x(\bold k)
+ (n_B(\omega_{\bold{k}})+n_F(\epsilon_{t,\bold{p} }))\,  \varphi_h(\bold p-\bold k)\bigg),
\end{align}

By substituting Eq.~\eqref{eq:varphit} back into Eqs.~\eqref{eq:eqnsforvarphi}, we find the following expression for $\varphi_x$ and $\varphi_h$: 
\begin{flalign}
\varphi_h(\bold p ) &= - \bigg[\mathcal M^h+\Upsilon^h - \mathcal N^h\Big(\mathcal  M^x+\Upsilon^x\Big)^{-1}\mathcal N^x\bigg]^{-1}_{\bold p ,\bold q}
\mathcal K^h_{ \bold q },\\
\label{eq:varphix}
\varphi_x(\bold k) &=\bigg[\mathcal{M}^x+\Upsilon^x- \mathcal{N}^x\Big(\mathcal M^h+\Upsilon^h\Big)^{-1}\mathcal N^h\bigg]^{-1}_{\bold k , \bold q }\,
\bigg(\mathcal N_x \Big( \mathcal M_h +\Upsilon_h\Big)^{-1}\mathcal K^h\bigg)_{ \bold q },
\end{flalign}
where repeated indices imply implicit summation, and for brevity we defined the following matrices:
\begin{subequations}
 \label{eq:M44all}
\begin{flalign}
\mathcal M^x_{\mb k , \mb q}
&=\delta_{\mb k , \mb q}\frac{2\pi}{A} g^2
\sum_{\mb p}\delta(\omega_{\mb k }-\epsilon_{t, \mb p + \mb k }+ \epsilon_{h,\mb p}) \left(n_F(\epsilon_{h,\mb p})-n_F(\epsilon_{t,\mb k+ \mb p })\right)
- \frac{4\pi^2}{A^2}g^4\sum_{\mb p} \delta(\omega_{\mb k }-\epsilon_{t, \mb p + \mb k }+ \epsilon_{h,\mb p}) \mc Q^x_{\mb q,\mb p,\mb k},\\
 \mathcal N^x_{\mb k , \mb q}
 &=\frac{2\pi}{A} g^2\delta(\omega_{\mb k}-\epsilon_{t, \mb q + \mb k}+\epsilon_{h, \mb q})  \big(n_F(\epsilon_{t,\mb q + \mb k})+n_B(\omega_{\mb k})\big)
 -\frac{4 \pi^2}{A^2}g^4 \sum_{\mb p}\delta(\omega_{\mb k}-\epsilon_{t, \mb p + \mb k}+\epsilon_{h, \mb p}) \mc Q^h_{\mb q,\mb p,\mb k}, \\
 \label{eq:Mh44specific}
\mathcal M^h_{\mb p,\mb q}
&= \delta_{\mb p, \mb q} \frac{2\pi}{A}g^2\sum_{\mb k }\delta(\omega_{\mb k}+\epsilon_{h, \mb p}-\epsilon_{t, \mb p+\mb k})
\left( n_F(\epsilon_{t, \mb p +\mb k }) + n_B(\omega_{\mb k })\right )
- \frac{4\pi^2}{A^2}g^4\sum_{\mb k }\delta(\omega_{\mb k}+\epsilon_{h, \mb p}-\epsilon_{t, \mb p+\mb k}) \mc Q^h_{\mb q,\mb p,\mb k},\\
\mathcal N^h_{ \mb p, \mb q} 
 &=\frac{2\pi}{A} g^2 \delta(\omega_{\mb q }+\epsilon_{h, \mb p }- \epsilon_{t, \mb p + \mb q }) \big(n_F(\epsilon_{h, \mb p}) - n_F(\epsilon_{t, \mb p+\mb q })\big)
 -\frac{4\pi ^2}{A^2}g^4\sum_{\mb k } \delta(\omega_{\mb k }-\epsilon_{t, \mb p + \mb k }+ \epsilon_{h,\mb p})\mc Q^x_{\mb q,\mb p,\mb k} ,
\end{flalign}
\end{subequations}
\begin{subequations}
\label{eq:M45all}
\begin{flalign}
  \mc Q^x_{\mb q,\mb p,\mb k}&=  
  (\Upsilon^t+\mathcal M^t)^{-1}_{\mb p +\mb k,\mb p +\mb k}
  (1+n_B(\omega_{\mb k})-n_F(\epsilon_{h, \mb p}))\delta(\omega_{\mb q}+\epsilon_{h, \mb p + \mb k -\mb q}-\epsilon_{t, \mb p + \mb k }) \big( n_F(\epsilon_{h,\mb p+\mb k-\mb q })- 
 n_F(\epsilon_{t,\mb p+\mb k}) \big ),\\
\mc Q^h_{\mb q,\mb p,\mb k}&=  
(\Upsilon^t +\mathcal M^t)^{-1}_{\mb p +\mb k,\mb p +\mb k}
(1+n_B(\omega_{\mb k})-n_F(\epsilon_{h,\mb p })) \delta(\omega_{ \mb k +\mb p -\mb q}+\epsilon_{h, \mb q}-\epsilon_{t,\mb p + \mb k})(n_B(\omega_{\mb k +\mb p -\mb q})+ n_F(\epsilon_{t, \mb p + \mb k })).
\end{flalign}
\end{subequations}
\end{widetext}

In Eqs.~\eqref{eq:M44all} we keep corrections which, although formally of order $O(g^4)$, may become relevant close to resonance -- see discussion in the next section.

We then obtain the conductivities Eq.~\eqref{eq:conductivitiesfromphi} by a numerical evaluation of Eqs.~\eqref{eq:varphit}--\eqref{eq:varphix} and their building blocks Eqs.~\eqref{eq:M44all},\eqref{eq:M45all},
with the choice $\tau_{i,\mb q}=\tau_0$ for all $i \in \{h,x,t\}$.

\begin{figure}
    \centering\includegraphics[width=0.89\linewidth]{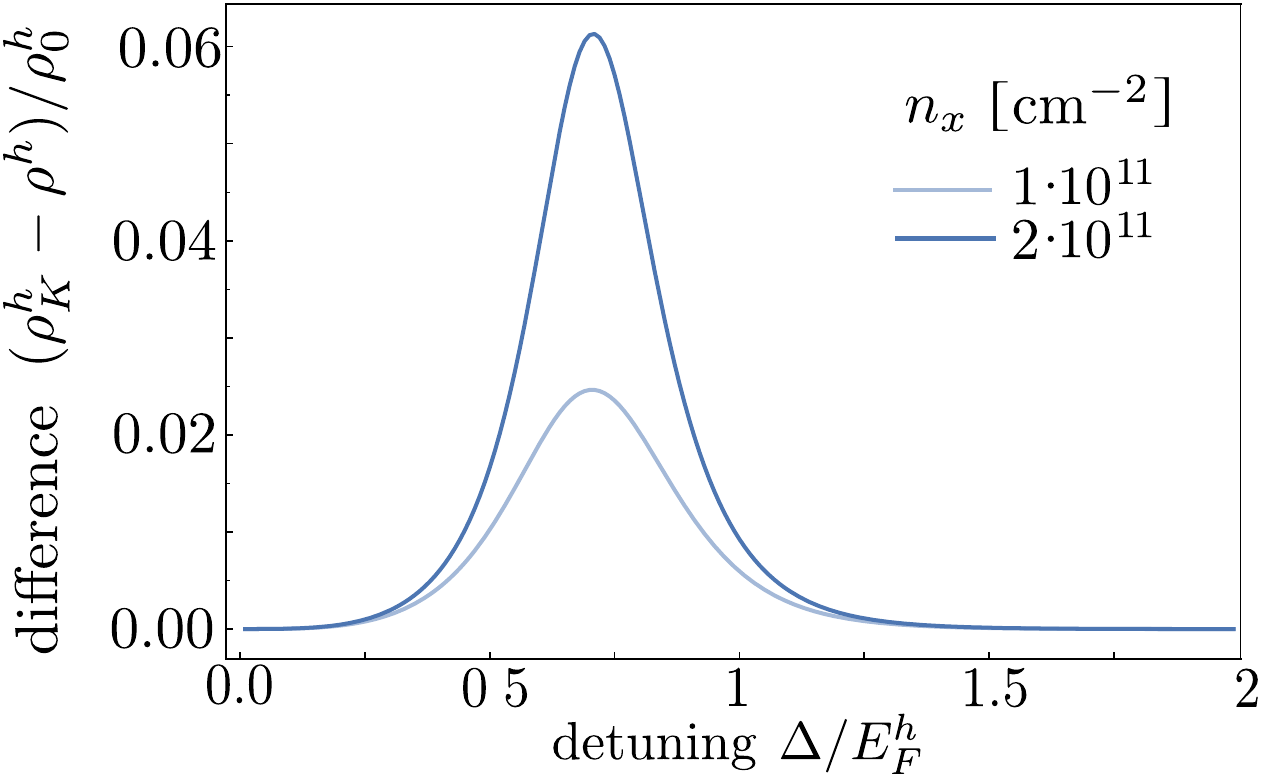}
    \caption{\textbf{Comparison of Kubo and Boltzmann approaches for the hole resistivity.} Difference of the hole resistivities as a function of the detuning $\Delta$ as obtained from Kubo's formula to leading order, $\rho^h_K$, and Boltzmann's equations as described in the previous appendix, $\rho^h$, for $t=1$ meV, $T =5.8$ K and for different values of the exciton density $n_x$. The relative difference arising from the $\mc O(g^4)$ corrections included in the kinetic theory treatment, is of a few percent of the background resistivity $\rho_0^h$.}
    \label{fig:5}
\end{figure}
\section{Linear response from Kubo's formula}\label{App:B}

Since the system hosts a small parameter, we can also analyze the conductivity by evaluating Kubo's formula
\begin{equation}
    \sigma^h_{ab}(\mathbf{q},\omega) = \frac{ie^2}{\omega} \left[ \chi^R_{\mathbf{j}_a\mathbf{j}_b}(\mathbf{q},\omega) + \delta_{ab} \frac{\langle n_h\rangle}{m_h} \right],
    \label{eq:cond}
\end{equation}
where $\chi^R_{\mathbf{j}_a\mathbf{j}_b}(\mathbf{q},\omega)$ is the retarded current-current response function, $n_h$ is the hole density, and $a,b\in\{x,y\}$. To lowest order in $g$, the diagrammatic expression for $\chi^R_{\mathbf{j}_a\mathbf{j}_b}(\mathbf{q},\omega)$ yields
\begin{fmffile}{currentcurrent}
	\begin{align}
\chi_{\mathbf j_a \mathbf j_b}^R(\mathbf{q},\omega) = 	
 \begin{gathered}
	\begin{fmfgraph}(70,15)
	\fmfleft{i}
	\fmfright{o}
	\fmf{phantom}{i,v1,v2,o}
 	\fmf{dashes,fore=(0.1,,0.1,,0.1)}{i,v1}
  \fmf{dashes}{v2,o}
	\fmffreeze
 \fmf{fermion,fore=(0.15,,0.2,,0.4),width=1.8,tension=1,left=0.7}{v2,v1}
 \fmf{fermion,fore=(0.15,,0.2,,0.4),width=1.8,left=0.7}{v1,v2}
    \fmfdot{v1,v2}
	\end{fmfgraph}	
	\end{gathered} + \mathcal{O}(g^4),
	\end{align}
\end{fmffile}
where the thick blue lines denote hole propagators, which are obtained by solving Dyson's equation to order $\mathcal{O}(g^2)$:
\begin{fmffile}{selfeng}
	\begin{align}
	\begin{gathered}
	\begin{fmfgraph*}(30,25)
	\fmfleft{i}
	\fmfright{o}
	\fmf{fermion,fore=(0.15,,0.2,,0.4),width=1.8}{i,o}
	\end{fmfgraph*}
	\end{gathered}=
	\begin{gathered}
	\begin{fmfgraph*}(30,20)
	\fmfleft{i}
	\fmfright{o}
	\fmf{fermion,fore=(0.3,,0.4,,0.8)}{i,o}
	\end{fmfgraph*}
	\end{gathered}+
	\begin{gathered}
	\begin{fmfgraph}(80,15)
	\fmfleft{i}
	\fmfright{o}
	\fmf{phantom}{i,v1,v3,o}
	\fmf{fermion,fore=(0.3,,0.4,,0.8),tension=5.}{i,v1}
	\fmf{fermion,fore=(0.15,,0.2,,0.4),tension=5,width=1.8}{v3,o}
 \fmf{boson,tension=1}{v3,v1}
 \fmf{phantom_arrow}{v3,v1}
 \fmf{dbl_plain_arrow,fore=(0.9,,0.3,,0.4),left=0.75}{v1,v3}
	\end{fmfgraph}	
	\end{gathered}.
 \label{eq:hole_selfeng}
	\end{align}
\end{fmffile}
In Eq.~\eqref{eq:hole_selfeng}, light-blue, red and squiggly lines denote bare-hole, trion and exciton propagators:
\begin{equation}
\begin{fmffile}{propagators}
    \begin{aligned}
    \begin{gathered}
	\begin{fmfgraph*}(30,20)
	\fmfleft{i}
	\fmfright{o}
	\fmf{fermion,fore=(0.3,,0.4,,0.8)}{i,o}
	\end{fmfgraph*}
	\end{gathered} 
 &= \frac{1}{\omega - \xi^h(\mathbf{p}) + i/2\tau^0_{h,\mathbf{q}}}  \\
    \begin{gathered}
	\begin{fmfgraph*}(30,20)
	\fmfleft{i}
	\fmfright{o}
	\fmf{dbl_plain_arrow,fore=(0.9,,0.3,,0.4)}{i,o}
	\end{fmfgraph*}
	\end{gathered} 
 &= \frac{1}{\omega - \xi^t(\mathbf{p}) + i/2\tau^0_{t,\mathbf{q}}}  \\
  \begin{gathered}
	\begin{fmfgraph*}(30,20)
	\fmfleft{i}
	\fmfright{o}
	\fmf{boson}{i,o}
    \fmf{phantom_arrow}{i,o}
	\end{fmfgraph*}
	\end{gathered} 
 &= \frac{1}{\omega - \omega(\mathbf{k})+  i/2\tau^0_{x,\mathbf{q}}},
    \end{aligned}
\end{fmffile}
\end{equation}
where we have modelled background scattering by adding a finite line broadening $1/2\tau_{i,\mathbf{q}} = 1/2\tau_0$.
Evaluating these diagrams and neglecting vertex corrections, one obtains the standard result
\begin{equation}
\begin{aligned}
&\lim_{\omega \rightarrow 0} \frac{1}{\omega} \left(\chi_{\mb j_x \mb j_x}^R(0,\omega) + \frac{\langle n_h \rangle}{m_h} \right) \\
&\qquad = \frac{i}{4\pi}\int_{0}^\infty\!\!\! dE E \int \frac{dz}{2\pi } \frac{ \mathcal A(k(E),z) }{\rm{Im} \Sigma^R_h(k(E),z)} n_F'(z),
\label{eq:response}
\end{aligned}
\end{equation}
where $k(E)^2 = 2m_h E$.
The current-current response function can be expressed directly in terms of the hole spectral function
\begin{equation}
   \mathcal A(\mathbf{k},z)= -2 \rm{Im} G^R(\mathbf{k},z)
\end{equation} and their perturbatively evaluated self energy
\begin{widetext}
\begin{equation}
\begin{aligned}
\Sigma^{R}_h(\mathbf{k},\omega)
    &= -g^2\int \frac{d^2 \mathbf q}{(2\pi)^2}\frac{ n_F(\epsilon_{t,\mathbf{q+k}}) + n_B(\omega_{\mathbf{q}})}{\epsilon_{t,\mathbf{q+k}} - i/2\tau_{t,\mb{q+k}} - \omega  - \omega_{\mathbf{q}} - i/2\tau_{x,\mb q}}  - \frac{i}{2\tau_{h,\mb k}} + \mathcal{O}(g^4);
  \end{aligned}
  \label{eq:SigmaRkubo}
\end{equation}
\end{widetext}
see Eq.~\eqref{eq:hole_selfeng}. By combining Eq.~\eqref{eq:cond} and Eq.~\eqref{eq:response}, we can express the conductivity in a standard form
\begin{equation}\label{eq:Kubo54}
\sigma^{h}_{xx}(\omega=0) \frac{h}{e^2}= \int_{-\mu}^\infty\!\!\! d\xi \left(\xi + \mu\right) n_F'(\xi)  \tau_{\mathrm{tot}}(\xi),
\end{equation}
where $\tau_{\mathrm{tot}}^{-1}(\xi) = -2\rm{Im} \Sigma^R_h(k(\xi+\mu),\xi)$.

The linear response result Eq.~\eqref{eq:Kubo54} coincides with that obtained by the kinetic method in the previous subsection, to the leading order $\mc O(g^2)$. 
Indeed, the term of order $g^2$ in $\mathcal{M}^h$ in Eq.~\eqref{eq:Mh44specific}, responsible for the most relevant many-body contribution to the hole resistivity, 
is identical to $2 \rm{Im} \Sigma^R_h(\mb p, \epsilon_{\mb p, h})$ in Eq.~\eqref{eq:SigmaRkubo} in the limit $1/2\tau_{i,\mb q}\rightarrow 0^+$. The Boltzmann approach also takes into account higher orders in $g^2$, however we checked numerically that $\mc O(g^4)$ corrections to the hole resistivity, albeit finite, are small for the considered parameters, see Fig.~\ref{fig:5}.

On the kinetic theory side, such formally $\mc O(g^4)$ corrections arise from the rightmost terms in Eqs.~\eqref{eq:M44all}, and especially in the hole resistivity case from 
$Q(\mb p, \mb q) \equiv - \frac{4\pi^2}{A^2}g^4\sum_{\mb k }\delta(\omega_{\mb k}+\epsilon_{h, \mb p}-\epsilon_{t, \mb p+\mb k}) \mc Q^h_{\mb q,\mb p,\mb k}$ in Eq.~\eqref{eq:Mh44specific}.
This correction becomes relevant for $\mb p \approx \mb q $ at small temperatures $k_BT \ll \hbar^2 n_x/m_x$, where the population of excitons at small momenta $\bold k \approx \bs 0$ is a sizeable fraction of $n_x$. In this regime, for small values of disorder, it can be that $\Upsilon^t+\mathcal{M}^t$ is dominated by $\mathcal{M}^t \propto g^2 n_x$.
This in turn implies that $Q(\mb q, \mb q) \sim g^2 n_x$, in contrast to its formal magnitude $\mc O(g^4)$. This provides a relevant contribution to the hole resistivity obtained from the kinetic theory method, which in the formalism of Kubo's equation would correspond to higher-order diagrammatic corrections to the bare conductivity bubble calculated in Eq.~\eqref{eq:response}.

\section{Trions}

\begin{figure}
    \centering
    \includegraphics[width=\linewidth]{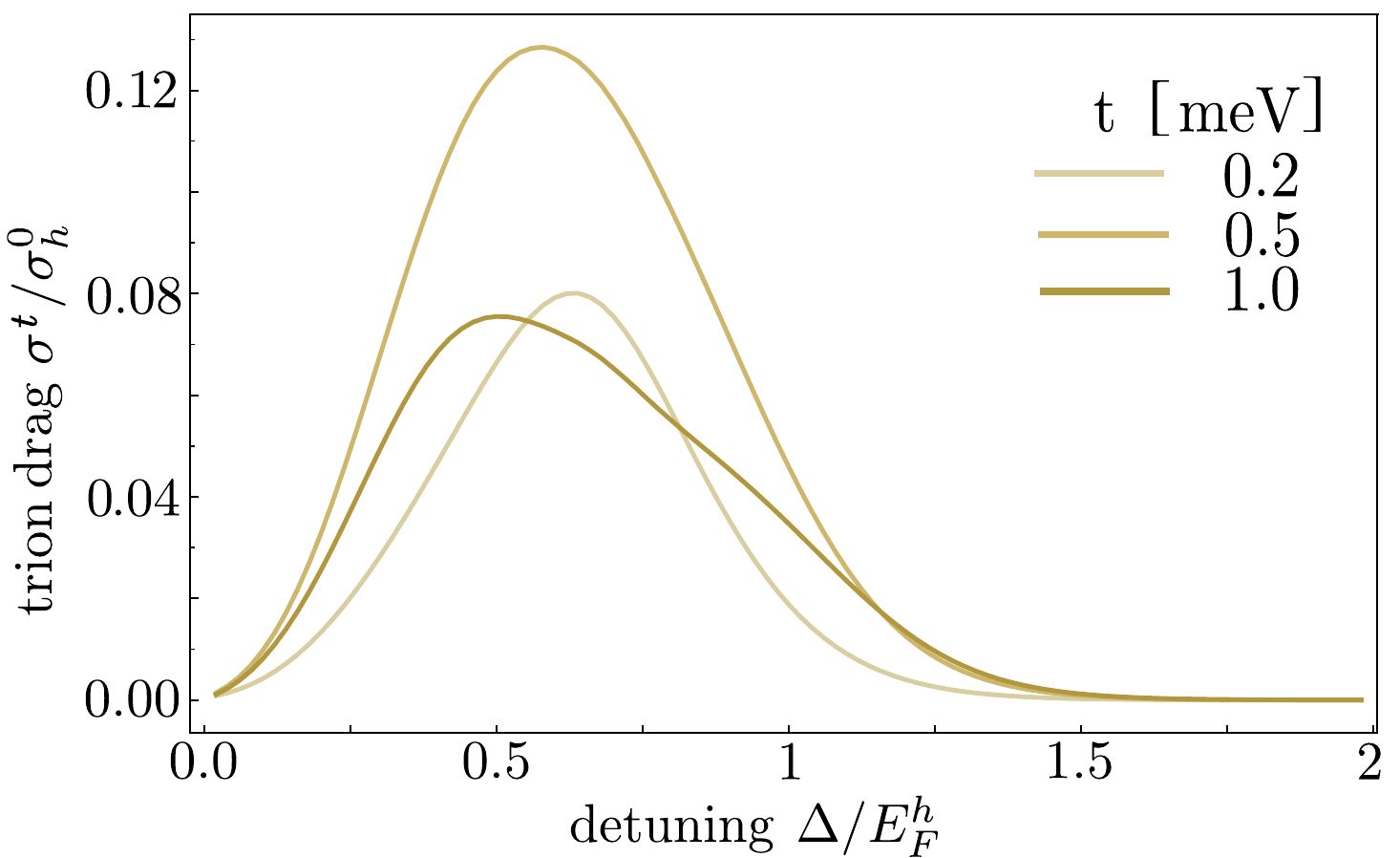}
    \caption{\textbf{Trion drag across resonance}. Trion conductivity as a function of the detuning $\Delta$, for $T=5.8$ K and different values of the tunneling parameter t.}
    \label{fig:6}
\end{figure}

The trions exhibit a drag effect similarly to the excitons.
The resulting rion conductivity depends strongly on resonant interactions, see Fig.~\ref{fig:6}. For small values of the interaction parameter $g \propto \rm t$ and small temperature, $\sigma^t$ reaches a maximum in the vicinity of the resonance $(\Delta \approx \Delta_\star)$ and increases with increasing $g$, as expected from the general discussion of hole-exciton-trion scattering given in the main text. At larger $g$, where $\mc M_i \gtrsim \Upsilon_i$ for $i\in\{h,x,t\}$, many-body corrections to the relaxation times become important. As a result, with increasing $g$, $\sigma^h$ decreases, $\sigma^x$ saturates, and $\sigma^t$ decreases before it saturates. Such corrections at large $g$ also entail less evident signatures nearby resonance, especially noticeable in the trion case. 
Remarkably, we find that the trion conductivity takes values of the same order of magnitude as the exciton conductivity. Therefore, the individual conductivities can be resolved by separately measuring charge currents in both the middle and the lower layer.

\section{Hydrodynamic model for ac conductivities}

The three-fluid model of the main text can be solved analytically.
The ac conductivities of the three particle species are obtained as:
\begin{align}
\label{eq:sigmaiomega}
\sigma^i(\Omega) &=\frac{{\sf A}_i} {1/\tau_i +i\Omega+ {\sf B}_i} \frac{n_i}{n_h} \sigma^h(\Omega),\\
\label{eq:sigmahomega}
\sigma^h(\Omega) &=\frac{e^2 n_h /m_h}{1/\tau_h+i\Omega+ \sum_{i=x,t} {\sf C}_i (1-\frac{{\sf A}_i}{i\Omega+1/\tau_{i}+{\sf B}_i})} ,
\end{align}
where we defined 
 \begin{subequations}
\begin{align}
&{\sf A}_i = \alpha_{ih} \frac{ n_h} {m_i}+\frac{\alpha_{\overline ih}\alpha_{tx} n_{\overline i} n_h / m_i }{\alpha_{\overline ih} n_h +\alpha_{tx} n_i + i\Omega +1/\tau_{\overline i} },\\
&{\sf B}_i =  \frac{\alpha_{tx}n_{\overline i}}{m_i}+\frac{\alpha_{ih} n_h}{m_i}-  \frac{\alpha_{tx}^2 n_x n_t/ m_i}{\alpha_{\overline ih} n_h+\alpha_{tx}n_i+ i\Omega +1/\tau_{\overline i}},\\
&{\sf C}_i= \alpha_{ih}n_i/m_h 
\end{align}
 \end{subequations}
and use the shorthand notation $i=x,t \leftrightarrow \overline i=t,x$ respectively.

We use Eqs.~\eqref{eq:sigmaiomega},\eqref{eq:sigmahomega} to fit the ac conductivities, obtained by solving the system of coupled Boltzmann's equations as described in the first appendix, reported in the main text. We find the following values for the fitting parameters: $\tau_h=9.4$ ps, $\tau_x=10.0 $ ps, $\tau_t=2.9$ ps, $\alpha_{th}=0.64 \frac{m_0}{n_0 \tau_0}$, $\alpha_{xh}=-0.48 \frac{m_0}{n_0 \tau_0}$, $\alpha_{tx}= 0.50  \frac{m_0}{n_0 \tau_0}$, where $n_0 =10^{12} \text{cm}^{-2}$, $m_0 = 0.25 m_{\sf e}$, and $\tau_0=10$ ps.

\end{document}